\documentstyle[12pt]{article}

\begin{document}

\begin{center}
Stabilization of highly dimensional statistical systems: Girko ensemble
\end{center}
\begin{center}
Maciej M. Duras
\end{center}
\begin{center}
Institute of Physics, Cracow University of Technology, 
ulica Podchor\c{a}\.zych 1, PL-30084 Cracow, Poland
\end{center}

\begin{center}
Email: mduras @ riad.usk.pk.edu.pl
\end{center}

\begin{center}
"Symposium on Entropy"; 
June 25th, 2000 - June 28th, 2000; 
Max Planck Institute for the Physics of Complex Systems;
Dresden, Germany (2000).
\end{center}

\begin{center}
AD 2000 June 21
\end{center}

\section{Abstract}
\label{sec-Abstract}
A quantum statistical system with energy dissipation is studied.
Its statisitics is governed by random complex-valued
non-Hermitean Hamiltonians belonging to complex Ginibre ensemble.
The eigenenergies are shown to form stable structure
in thermodynamical limit (large matrix dimension limit).
Analogy of Wigner and Dyson with system of electrical charges
is drawn.

\section{Summary}
\label{sec-Summary}
A complex quantum system with energy dissipation is considered.
The quantum Hamiltonians $H$ belong the complex Ginibre ensemble.
The complex-valued eigenenergies $Z_{i}$
are random variables.
The second differences $\Delta^{1} Z_{i}$ are also
complex-valued random variables.
The second differences have their
real and imaginary parts and also 
radii (moduli) and main arguments (angles).
For $N$=3 dimensional Ginibre ensemble
the distributions of above random variables are provided
whereas for generic $N$- dimensional Ginibre ensemble 
second difference distribution is
analytically calculated.
The law of homogenization of eigenergies is formulated.
The analogy of Wigner and Dyson
of Coulomb gas of electric charges is studied.

\section{Introduction}
\label{sec-introduction}
We study generic quantum statistical systems with energy dissipation.
The quantum Hamiltonian operator $H$ is in given
basis of Hilbert's space a matrix with random
elements $H_{ij}$
\cite{Haake 1990,Guhr 1998,Mehta 1990 0}.
The Hamiltonian $H$ is not hermitean operator, thus
its eigenenergies $Z_{i}$ are
complex-valued random variables.
We assume that distribution of $H_{ij}$
is governed by Ginibre ensemble
\cite{Haake 1990,Guhr 1998,Ginibre 1965,Mehta 1990 1}.
$H$ belongs to general linear Lie group GL($N$, {\bf C}),
where $N$ is dimension and {\bf C} is complex numbers field.
Since $H$ is not hermitean, therefore quantum system is
dissipative system. Ginibre ensemble of random matrices
is one of many 
Gaussian Random Matrix ensembles GRME.
The above approach is an example of Random Matrix theory RMT
\cite{Haake 1990,Guhr 1998,Mehta 1990 0}.
The other RMT ensembles are for example
Gaussian orthogonal ensemble GOE, unitary GUE, symplectic GSE,
as well as circular ensembles: orthogonal COE,
unitary CUE, and symplectic CSE.
The distributions of the eigenenergies 
$Z_{1}, ..., Z_{N}$
for $N \times N$ Hamiltonian matrices is given by Jean Ginibre's formula 
\cite{Haake 1990,Guhr 1998,Ginibre 1965,Mehta 1990 1}:
\begin{eqnarray}
& & P(z_{1}, ..., z_{N})=
\label{Ginibre-joint-pdf-eigenvalues} \\
& & =\prod _{j=1}^{N} \frac{1}{\pi \cdot j!} \cdot
\prod _{i<j}^{N} \vert z_{i} - z_{j} \vert^{2} \cdot
\exp (- \sum _{j=1}^{N} \vert z_{j}\vert^{2}),
\nonumber
\end{eqnarray}
where $z_{i}$ are complex-valued sample points
($z_{i} \in {\bf C}$).
For Ginibre ensemble we define complex-valued spacings
$\Delta^{1} Z_{i}$ and second differences $\Delta^{2} Z_{i}$:
\begin{equation}
\Delta^{1} Z_{i}=Z_{i+1}-Z_{i}, i=1, ..., (N-1),
\label{first-diff-def}
\end{equation}
\begin{equation}
\Delta ^{2} Z_{i}=Z_{i+2} - 2Z_{i+1} + Z_{i}, i=1, ..., (N-2).
\label{Ginibre-second-difference-def}
\end{equation}
The $\Delta^{2} Z_{i}$ are extensions of
real-valued second differences
\begin{equation}
\Delta^{2} E_{i}=E_{i+2}-2E_{i+1}+E_{i}, i=1, ..., (N-2),
\label{second-diff-def}
\end{equation}
of adjacent ordered increasingly real-valued energies $E_{i}$
defined for
GOE, GUE, GSE, and Poisson ensemble PE
(where Poisson ensemble is composed of uncorrelated
randomly distributed eigenenergies)
\cite{Duras 1996 PRE,Duras 1996 thesis,Duras 1999 Phys,Duras 1999 Nap,Duras 1996 APPB,Duras 1997 APPB}.

There is an analogy of Coulomb gas of 
unit electric charges pointed out by Eugene Wigner and Freeman Dyson.
A Coulomb gas of $N$ unit charges moving on complex plane (Gauss's plane)
{\bf C} is considered. The vectors of positions
of charges are $z_{i}$ and potential energy of the system is:
\begin{equation}
U(z_{1}, ...,z_{N})=
- \sum_{i<j} \ln \vert z_{i} - z_{j} \vert
+ \frac{1}{2} \sum_{i} \vert z_{i}^{2} \vert. 
\label{Coulomb-potential-energy}
\end{equation}
If gas is in thermodynamical equilibrium at temperature
$T= \frac{1}{2 k_{B}}$ 
($\beta= \frac{1}{k_{B}T}=2$, $k_{B}$ is Boltzmann's constant),
then probability density function of vectors of positions is 
$P(z_{1}, ..., z_{N})$ Eq. (\ref{Ginibre-joint-pdf-eigenvalues}).
Complex eigenenergies $Z_{i}$ of quantum system 
are analogous to vectors of positions of charges of Coulomb gas.
Moreover, complex-valued spacings $\Delta^{1} Z_{i}$
are analogous to vectors of relative positions of electric charges.
Finally, complex-valued
second differences $\Delta^{2} Z_{i}$ 
are analogous to
vectors of relative positions of vectors
of relative positions of electric charges.

The $\Delta ^{2} Z_{i}$ have their real parts
${\rm Re} \Delta ^{2} Z_{i}$,
and imaginary parts
${\rm Im} \Delta ^{2} Z_{i}$, 
as well as radii (moduli)
$\vert \Delta ^{2} Z_{i} \vert$,
and main arguments (angles) ${\rm Arg} \Delta ^{2} Z_{i}$.

\section{Second Difference Distributions}
\label{sec-second-difference-pdf}
We define following random variables
for $N$=3 dimensional Ginibre ensemble:
\begin{equation}
Y_{1}=\Delta ^{2} Z_{1}, 
A_{1}= {\rm Re} Y_{1}, B_{1}= {\rm Im} Y_{1},
\label{Ginibre-Y1A1B1-def}
\end{equation}
\begin{equation}
R_{1} = \vert Y_{1} \vert, \Phi_{1}= {\rm Arg} Y_{1},
\label{Ginibre-polar-second-diff-def}
\end{equation}
and for the generic $N$-dimensional Ginibre ensemble
\cite{Duras 2000 JOptB}:
$W_{1}=\Delta ^{2} Z_{1}$.

Their distributions for 
are given by following formulae 
\cite{Duras 2000 JOptB}:
\begin{eqnarray}
& & f_{Y_{1}}(y_{1})=f_{(A_{1}, B_{1})}(a_{1}, b_{1})=
\label{Ginibre-marginal-pdf-Y1-def} \\
& & =\frac{1}{576 \pi} [ (a_{1}^{2} + b_{1}^{2})^{2} + 24]
\cdot \exp (- \frac{1}{6} (a_{1}^{2}+a_{2}^{2})).
\nonumber
\end{eqnarray}
\begin{equation}
f_{A_{1}}(a_{1})=
\frac{\sqrt{6}}{576 \sqrt{\pi}} (a_{1}^{4}+6a_{1}^{2}+ 51)
\cdot \exp (- \frac{1}{6} a_{1}^{2}),
\label{Ginibre-marginal-pdf-Y1Re-def}
\end{equation}
\begin{equation}
f_{B_{1}}(b_{1})=
\frac{\sqrt{6}}{576 \sqrt{\pi}} (b_{1}^{4}+6b_{1}^{2}+ 51)
\cdot \exp (- \frac{1}{6} b_{1}^{2}),
\label{Ginibre-marginal-pdf-Y1Im-def}
\end{equation}
\begin{eqnarray}
& & f_{R_{1}}(r_{1})=
\label{Ginibre-polar-second-diff-result} \\
& & \Theta(r_{1}) \frac{1}{288}r_{1}(r_{1}^{4}+24) \cdot \exp(- \frac{1}{6} r_{1}^{2}),
\nonumber \\
& & f_{\Phi_{1}}(\phi_{1})= \frac{1}{2 \pi}, \phi_{1} \in [0, 2 \pi].
\nonumber
\end{eqnarray}
\begin{eqnarray}
& & P_{3}(w_{1})=
\label{W1-pdf-I-result} \\
& & = \pi^{-3} \sum_{j_{1}=0}^{N-1} \sum_{j_{2}=0}^{N-1} \sum_{j_{3}=0}^{N-1}
\frac{1}{j_{1}!j_{2}!j_{3}!}I_{j_{1}j_{2}j_{3}}(w_{1}),
\nonumber \\
& & I_{j_{1}j_{2}j_{3}}(w_{1})=
\label{W1-pdf-I-F} \\
& & = 2^{-2j_{2}} 
\frac{\partial^{j_{1}+j_{2}+j_{3}}}
{\partial^{j_{1}} \lambda_{1} \partial^{j_{2}} \lambda_{2}
\partial^{j_{3}} \lambda_{3}}
F(w_{1},\lambda_{1},\lambda_{2},\lambda_{3}) \vert _{\lambda_{i}=0},
\nonumber
\end{eqnarray}
\begin{eqnarray}
& & F(w_{1},\lambda_{1},\lambda_{2},\lambda_{3})=
\label{W1-pdf-I-F-final} \\
& & = A(\lambda_{1},\lambda_{2},\lambda_{3})
\exp[-B(\lambda_{1},\lambda_{2},\lambda_{3}) \vert w_{1} \vert^{2}],
\nonumber
\end{eqnarray}
\begin{eqnarray}
& & A(\lambda_{1},\lambda_{2},\lambda_{3})=
\label{W1-pdf-I-A} \\
& & =\frac{(2\pi)^{2}}
{(\lambda_{1}+\lambda_{2}-\frac{5}{4}) 
\cdot (\lambda_{1}+\lambda_{3}-\frac{5}{4})-(\lambda_{1}-1)^{2}},
\nonumber \\
& & B(\lambda_{1},\lambda_{2},\lambda_{3})=
\label{W1-pdf-I-B} \\
& & =(\lambda_{1}-1) \cdot \frac{2 \lambda_{1}-\lambda_{2}-\lambda_{3}+\frac{1}{2}}
{2 \lambda_{1}+\lambda_{2}+\lambda_{3}-\frac{9}{2}}.
\nonumber
\end{eqnarray}

\section{Conclusions}
\label{sect-conclusions}
We compare second difference distributions for different ensembles 
by defining following dimensionless second differences:
\begin{equation}
C_{\beta} = \frac{\Delta^{2} E_{1}}{<S_{\beta}>},
\label{rescaled-second-diff-GOE-GUE-GSE-PE}
\end{equation}
\begin{equation}
X_{1}=\frac{A_{1}}{<R_{1}>},
\label{Ginibre-X1-def} 
\end{equation}
where $<S_{\beta}>$ are
the mean values of spacings 
for GOE(3) ($\beta=1$),
for GUE(3) ($\beta=2$),
for GSE(3) ($\beta=4$), for PE ($\beta=0$)
\cite{Duras 1996 PRE,Duras 1996 thesis,Duras 1999 Phys,Duras 1999 Nap,Duras 1996 APPB,Duras 1997 APPB},
and $<R_{1}>$ is mean value of radius $R_{1}$ 
for $N$=3 dimensional Ginibre ensemble \cite{Duras 2000 JOptB}.

On the basis of comparison of results for
Gaussian ensembles, Poisson ensemble, and Ginibre ensemble
we formulate homogenization law
\cite{Duras 1996 PRE,Duras 1996 thesis,Duras 1999 Phys,Duras 1999 Nap,Duras 1996 APPB,Duras 1997 APPB,Duras 2000 JOptB}: 
{\it Eigenenergies for Gaussian ensembles, for Poisson ensemble,
and for Ginibre ensemble tend to be homogeneously distributed.}
The second differences' distributions assume global maxima at origin
for above ensembles.
For Coulomb gas 
the vectors of relative positions of vectors
of relative positions of charges statistically vanish.
It can be called stabilisation
of structure of system of electric charges. 

\section{Acknowledgements}
\label{sect-acknowledgements}
It is my pleasure to most deeply thank Professor Jakub Zakrzewski
for formulating the problem.

\end{document}